\documentclass[
 aip,
 amsmath,
 amssymb,
 reprint,
 sort&compress,
 numbers
]{revtex4-2}

\usepackage{graphicx} 
\usepackage{dcolumn} 
\usepackage{bm} 
\usepackage[utf8]{inputenc}
\usepackage[T1]{fontenc}
\usepackage{mathptmx}
\usepackage{tikz-cd}
\usepackage{etoolbox}
\usepackage{comment}
\usepackage{lineno}
\usepackage{color}
\usepackage{siunitx} 
\usepackage[version=4]{mhchem} 
\usepackage{hyperref} 
\usepackage[capitalise, nameinlink, noabbrev]{cleveref}
\usepackage{xcolor}
\usepackage{soul}
\hypersetup{
    colorlinks=true, 
    linkcolor=blue,
    filecolor=magenta,
    citecolor=blue,
    urlcolor=blue
} 

\DeclareUnicodeCharacter{2212}{-}
\DeclareUnicodeCharacter{2032}{'}

\preprint{AIP/123-QED}

\begin{document}
\title{Multi-Spectroscopic Method to Quantify Rapid Decomposition of an Organophosphate Simulant Using Reactive Materials as a Function of Metal Powder Chemistry and Temperature}

\author{Preetom Borah}
\email{pborah1@jhu.edu}
\affiliation{Department of Materials Science and Engineering, Johns Hopkins University, Baltimore, MD, 21218}
\affiliation{Hopkins Extreme Materials Institute, Johns Hopkins University, Baltimore, MD, 21218}

\author{Elif Irem Senyurt}
\affiliation{Department of Materials Science and Engineering, Johns Hopkins University, Baltimore, MD, 21218}
\affiliation{Hopkins Extreme Materials Institute, Johns Hopkins University, Baltimore, MD, 21218}
\affiliation{Department of Chemical and Materials Engineering, New Jersey Institute of Technology, Newark, NJ, 07102}

\author{Rohit Berlia}
\affiliation{Department of Materials Science and Engineering, Johns Hopkins University, Baltimore, MD, 21218}
\affiliation{Hopkins Extreme Materials Institute, Johns Hopkins University, Baltimore, MD, 21218}

\author{Jesse Grant}
\affiliation{Department of Materials Science and Engineering, Johns Hopkins University, Baltimore, MD, 21218}
\affiliation{Hopkins Extreme Materials Institute, Johns Hopkins University, Baltimore, MD, 21218}

\author{Edward Dreizin}
\affiliation{Department of Chemical and Materials Engineering, New Jersey Institute of Technology, Newark, NJ, 07102}

\author{Timothy P. Weihs}
\email{weihs@jhu.edu}
\affiliation{Department of Materials Science and Engineering, Johns Hopkins University, Baltimore, MD, 21218}
\affiliation{Hopkins Extreme Materials Institute, Johns Hopkins University, Baltimore, MD, 21218}

\begin{abstract}
The development of advanced diagnostic systems to measure and optimize emerging energetic material performance is critical for the defeat of Chemical Warfare Agents (CWAs). This study presents an integrated multi-spectroscopic approach to monitor the interaction between a CWA simulant, Diisopropyl Methyl Phosphonate (DIMP), and combusting composite metal particles. A custom benchtop Polygonal Rotating Mirror Infrared Spectrometer (PRiMIRS), equipped with a customizable experimental chamber, is employed to observe DIMP decomposition. Tunable Diode Laser Absorption Spectroscopy (TDLAS) is used to measure path-averaged gas temperature profiles during combustion. 

In the experiment, the chamber is preheated to evaporate liquid DIMP. Various composite metal powders – (Al-8Mg):3Zr, (Al-8Mg):Zr, 2(Al-8Mg):Zr, and 4(Al-8Mg):Zr – are placed on a stainless steel mount and ignited using 3Al-2Ni sputter-deposited nanolayered foils. The combusting metal particles mix with the DIMP vapor, initiating chemical and thermal interactions. PRiMIRS captures DIMP spectral evolution, while TDLAS simultaneously monitors gas temperature. 

A spectral defeat parameter was developed to enable quantitative real-time assessment of the DIMP destruction. It uses infrared light absorption by both from DIMP and its immediate decomposition products – Isopropyl Methyl Phosphonate (IMP) and Isopropyl Alcohol (IPA). Fourier Transform Infrared Spectroscopy (FTIR) serves as a secondary verification tool quantifying the decomposition products over extended timeframes, and Transmission Electron Microscopy (TEM) confirms the expected metal oxide dispersion within the reaction space. This study reports variability in DIMP defeat as a function of metal powder stoichiometry, metal powder loading, and path-averaged gas temperature profiles, offering critical insights into optimizing reactive materials for effective CWA neutralization.

\end{abstract}

\maketitle{}

\section{Introduction}
Chemical warfare agents (CWAs) have become an increasing concern due to their devastating use against civilian populations, as seen in Syria in 2018 \cite{2018SyrianPost,2018SyriaCNN} and Japan during 1994 and 1995 \cite{Yanagisawa2006SarinEffects}. Organophosphate nerve agents, in particular, disrupt acetylcholinesterase activity in the nervous system, causing severe and often fatal symptoms. These attacks have resulted in thousands of injuries and deaths  \cite{Yanagisawa2006SarinEffects}, underscoring the urgent need for effective countermeasures. To mitigate future threats, it is necessary to develop advanced materials and methods to rapidly neutralize CWAs and biological warfare agents (BWAs) before they can be deployed. One promising approach leverages energetic materials to thermally degrade these agents through rapid heating. For instance, the exothermic reactions of composite metal powders have demonstrated potential in neutralizing BWAs, thus placing them as strong candidates for countering CWAs as well \cite{Reverberi2023OrganicPerspectives}. Here we describe the development and methodology behind a Multi-spectroscopic System approach to quantify the decomposition of a CWA simulant via combusting composite reactive metal powders as a function of metal powder chemistry and measured gas temperature profiles. Composite reactive metal powders exhibit a wide variety of ignition and combustion characteristics that can be controlled by tuning their chemistry and morphology \cite{Wainwright2018ObservationsEnvironments,VummidiLakshman2019ThePowders,Stamatis2020CombustionParticles,Overdeep2015UsingNanolaminates,Yetter2009MetalNanotechnology}. To understand their interaction with CWA simulants, three decomposition pathways must be considered: (a) homogeneous global thermal defeat due to the elevated gas temperatures produced by combusting metal particles, (b) local thermal defeat of CWA simulants that are directly adjacent to the combusting metal powders, and (c) chemical defeat of CWA simulants due to molecular interaction with combustion products. Naturally, these factors will overlap with one another and it is difficult to distinguish one mode of decomposition from others in such a dynamic environment. Combustion of the metal powders will result in high local temperatures \cite{Sundaram2015TranslatedFrom,Millogo2020CombustionPowders,Glumac2005TemperatureDioxide,Das2025InteractionSurfaces} and oxides that are known to neutralize CWAs and their simulants \cite{Vasudevan2023RemovalSurfaces,Biswas2021High-TemperatureDynamics,Stengl2016NanostructuredAgents,Mukhopadhyay2021Vapor-phaseOxides,Ballow2020ZirconiumDecomposition,AdsorptionNTIS,Holdren2019AdsorptionOxide,Jeon2019ConformalAgents,Wang2021ThermalClusters,Kim2023MetalIrradiation,Denchy2021AdsorptionClusters,Tesvara2023DecompositionTiO2110}.  The combustion behavior of metals varies significantly depending on whether they burn in the gas phase or the condensed phase, leading to substantial differences in the size and number density of metal oxides within a combustion plume. Metals like Al and Mg, which combust in the vapor phase, generate a high density of nanoscale oxides, often visible as soot or smoke in the plume. In contrast, metals such as Ti or Zr burn in the condensed phase, producing a lower density of microscale oxide particles. As a result, even when their local heat production is comparable, the ability of their oxide products to neutralize simulants may differ significantly. To optimize metal powders for simulant neutralization, advanced diagnostic systems are essential to monitor both temperature and the molecular evolution of CWAs at fast timescales ($\ge$100 Hz) and over short durations ($<5$s).  The decomposition of CWA’s and CWA simulants such as diisopropyl methylphosphonate (DIMP), dimethyl methylphosphonate (DMMP), dimethyl nitrophenyl phosphate (DMNP), and triethyl phosphate (TEP) have been studied using both experimental and computational methods \cite{Zegers1998Gas-phaseMethylphosphonate,Yuan2019T-jumpMethylphosphonate,Rahman2022DirectAbsorption,Mott2012CalculatedSimulants,Neupane2019InfraredSimulants,Vasudevan2023RemovalSurfaces,Biswas2021High-TemperatureDynamics,Stengl2016NanostructuredAgents,Ballow2020ZirconiumDecomposition,AdsorptionNTIS,Holdren2019AdsorptionOxide,Jeon2019ConformalAgents,Wang2021ThermalClusters,Plonka2019EffectMOF-808,Kim2023MetalIrradiation,Jeon2020KineticsSpectroscopy,Denchy2021AdsorptionClusters,Tesvara2023DecompositionTiO2110,VanBuren2021CustomAgents,Nawaa2019ThermalAgents,Li2025InteractionSurfaces} , yet the decomposition pathways are complex and are sensitive to experimental design and measurement equipment. Zegers and Fisher et al \cite{Zegers1998Gas-phaseMethylphosphonate} identified a decomposition pathway for DIMP using a combination of Gas Chromatography - Mass Spectrometry (GC-MS) and Fourier transform infrared spectroscopy (FTIR) describing the formation of isopropyl methyl phosphonate (IMP), methyl phosphonic acid (MPA), propene, 2-propanol, and methyl(oxo) phosphoniumolate (MOPO):

\begin{equation}
\begin{tikzcd}
  \label{Eqn: DIMPpathway}
  && [-2em] MPA + Propene\\
    DIMP \arrow[r] & [-0em] IMP + Propene \arrow[dr] \arrow[ur]   \\
    &&[-3em] MOPO + 2-Propanol
\end{tikzcd}
\end{equation}


Subsequent studies have expanded on this pathway demonstrating variation at higher heating rates \cite{Yuan2019T-jumpMethylphosphonate} and in the presence of oxygen \cite{Senyurt2024ExperimentalDIMP} motivating further studies. While the GC-MS technique has high sensitivity to changes in molecular structure, the technique requires live analyte sampling and is not well-suited for measuring fast phenomena \cite{Zegers1998Gas-phaseMethylphosphonate,Yuan2019T-jumpMethylphosphonate,Rahman2022DirectAbsorption,Mukhopadhyay2021Vapor-phaseOxides}. Absorption spectroscopy techniques like Fourier transform infrared spectroscopy (FTIR) are popular alternatives for monitoring organic compounds, such as DIMP, TEP, and DMMP \cite{Neupane2018ShockCombustion,Neupane2020DMMPCO,Biswas2021High-TemperatureDynamics,Stengl2016NanostructuredAgents,Mukhopadhyay2021Vapor-phaseOxides}, as their functional groups absorb at different frequencies and absorption is linearly related to concentration, as shown by the Beer-Lambert law in \cref{Eqn: Beerslaw}.
\begin{equation}
\label{Eqn: Beerslaw}
\log \left( \frac{I_o}{I} \right) = A = \varepsilon \cdot l \cdot c
\end{equation}
where $A$ is the absorbance, $I$ is the transmission intensity, $I_o$ is the incident radiation, $\epsilon$ is the molar absorptivity coefficient, $l$ is the path length, and $c$ is the concentration of measured species. Another advantage to absorption spectroscopy is the ability to monitor experiments remotely without requiring live analyte sampling or post-reaction material analysis \cite{Zegers1998Gas-phaseMethylphosphonate,Yuan2019T-jumpMethylphosphonate,Neupane2019InfraredSimulants,Vasudevan2023RemovalSurfaces,Mukhopadhyay2021Vapor-phaseOxides}. FTIR offers high wavelength resolution and the ability to sample across a broad wavelength range, making it an effective tool for monitoring multiple functional groups within the IR spectrum. This capability allows for precise tracking of the evolution of initial reactants, intermediate species, and final byproducts. However, achieving such broad spectral coverage at high resolution results in time resolutions slower than 1 Hz, presenting a similar limitation to that of mass spectrometry (MS) for studying faster phenomena. 

Quantum Cascade Lasers (QCL’s) have also been used for absorption spectroscopy to monitor organo-phosphates at repetition rates well above kHz and approaching MHz \cite{Neupane2019InfraredSimulants,Neupane2020DMMPCO}. The drawback to QCL’s, however, is a very narrow cross section (0.1 cm$^{-1}$) that can only probe single functional groups. A narrow bandwidth can hinder the ability to monitor signal from simulants while measuring their byproducts, particularly when CWA simulants and their byproducts have overlapping peaks in the IR regime. 

External cavity quantum cascade lasers (ECQCLs) are an emerging technology capable of operating at scan rates exceeding a few hundred Hz, high resolution, and moderate spectral range. They have been used to monitor organic compounds in chemical environments but do not come without a substantial level of engineering complexity. The ability to design, build, and tailor these systems for unique experiments is challenging in conjunction with combustion environments but has begun recent implementation. \cite{Phillips2019CharacterizationSpectroscopy,Phillips2017StandoffLaser}.

A reliable and well validated alternative to the above methods is a rapid scanning dispersive spectrometer with kHz repetition rates and the ability to probe moderate wavelength ranges with more approachable optical designs \cite{Hand1970FlashMixtures,Ogawa1970ReactionSpectroscopy,Jensen1967KineticsSpectroscopy,Hexter1968InfraredSecond}. These systems are low cost and highly adaptable to meet specific experimental needs. A 2-slit system with a rotating grating was recently developed and used to characterize neutralization of DIMP \cite{Butler2023ASpectroscopy}. Building from this earlier success, we then developed a system that is referred to as a \textbf{P}olygonal \textbf{R}otat\textbf{i}ng\textbf{ M}irror \textbf{I}nfra\textbf{R}ed \textbf{S}pectrometer (PRiMIRS) \cite{Borah2024DevelopmentEnvironment}. This system was used to evaluate the first measurements of prompt defeat of a CWA simulant using reactive material in a non high-explosive (HE) environment. The scope is expanded in this study to begin to optimize RM chemistry to enhance CWA simulant decomposition. More robust details on the development of this system and its performance can be seen in Borah \textit{et al}. \cite{Borah2024DevelopmentEnvironment}. 

In addition to monitoring DIMP decomposition, it is essential to characterize properties specific to the combusting metal particles to better understand their role in the reaction environment. One key parameter is the rise in local gas temperature during particle combustion. Thermal measurements are typically conducted using either thermocouples or optical diagnostics. Although thermocouples offer high accuracy and precision, their inherently slow response times—particularly under rapidly changing conditions and sensitivity to harsh conditions—limit their usefulness in high-speed combustion environments, with effective sampling rates often below 1 Hz. \cite{Farahmand2001EXPERIMENTALAIR,Hashemian1990RESPONSETHERMOCOUPLES,Li2018AFurnace,Oliveira2022ThermocoupleProblems} Given the need to monitor DIMP evolution at timescales approaching 100 Hz, thermocouples are not well-suited for this application. 

Optical techniques for temperature measurement include infrared (IR) imaging and various spectroscopic methods. IR cameras detect radiation primarily emitted from surfaces and are typically calibrated for solid or liquid-phase temperatures, making them less effective for measuring temperature changes in gases within a sample cell. In contrast, emission spectroscopy is widely used in combustion research to probe the high temperatures directly associated with burning species. Temperature can be extracted from the intensity of emitted light, as demonstrated in \cite{Bonefacic2015TWO-COLORPHOTODIODES}, with additional examples using two-color and three-color pyrometry techniques provided in \cite{Panagiotou1996MeasurementsPyrometry,Goroshin2007EmissionSuspensions,Hossain2013Three-dimensionalTechniques,McNesby2021ImagingFilter}. These approaches are particularly well-suited for characterizing the localized high temperatures at or near combusting metal particles. 

However, our focus is on understanding how burning metal particles contribute to broader increases in background or global gas temperatures—an effect more relevant to the spatial distribution of DIMP vapor. Since the ambient gas does not emit detectable radiation under these conditions, we employ absorption spectroscopy. Specifically, we use tunable diode laser absorption spectroscopy (TDLAS), a technique shown to be effective for real-time gas-phase temperature measurements under similar experimental conditions by Mattison \textit{et al}. \cite{Mattison2006DEVELOPMENTBy,Mattison2003PulseSensors}, and followed up by Murzyn \textit{et al}.\cite{Murzyn2018HighSpectroscopy}. Our iteration of TDLAS is described in further detail in Borah \textit{et al}. \cite{Borah2024DevelopmentEnvironment}. 

This study presents the first known measurements demonstrating variable prompt defeat of CWAS as a function of metal powder chemsitry and gas temperature using combusting composite metal powders without the use of HE. Here we leverage the combined operation of PRiMIRS and TDLAS to explore whether reactive materials such as composite metal powders can promptly neutralize a CWA simulant and whether variations in powder chemistry and powder loading play a role in the level of decomposition. In doing so, we use PRiMIRS to monitor spectral profiles of DIMP and its byproducts and quantify the level of decomposition for a given rise in background temperature that is simultaneously quantified using TDLAS. Lastly, we use FTIR and TEM (Transmission Electron Microscopy) to confirm evidence of decomposition products as well as metal oxide products within the experimental space. We hypothesize that among formulations that range from condensed phase to vapor phase product generators, we will observe elevated defeat from vapor phase generators due to increased particle surface interaction with DIMP. 

\section{Materials and Methods}

\subsection{Experimental Chamber (DIMP Reactor)}

\begin{figure*}[!htbp]
    \centering
      \includegraphics[trim=5.5cm 9cm 5.5cm 7cm, clip=true,width=0.95\linewidth]{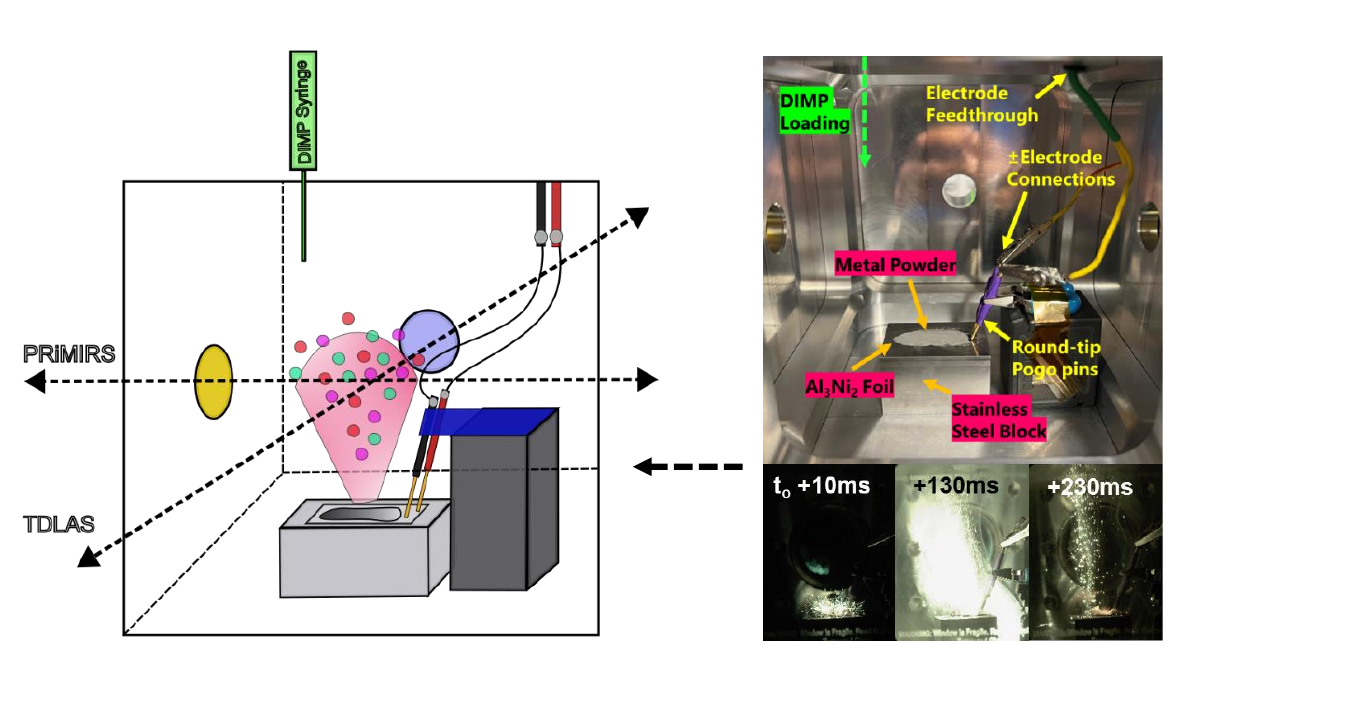}
      \caption{Modular Experimental Chamber: (a) Schematic showing how powders loaded on a reactive foil on top of a stainless steel block are ignited using 3Al-2Ni foil and subsequent combustion products propagate throughout the cell containing vaporized DIMP. (b) Photographs of the cell pre-ignition and snapshots after ignition (t$_0$) are shown to the right.}
    \label{fig:DIMPreactor}
\end{figure*}

\begin{figure*}[!htbp]
    \centering
    \includegraphics[trim=2cm 3cm 2cm 4cm, clip=true,width=0.85\linewidth]{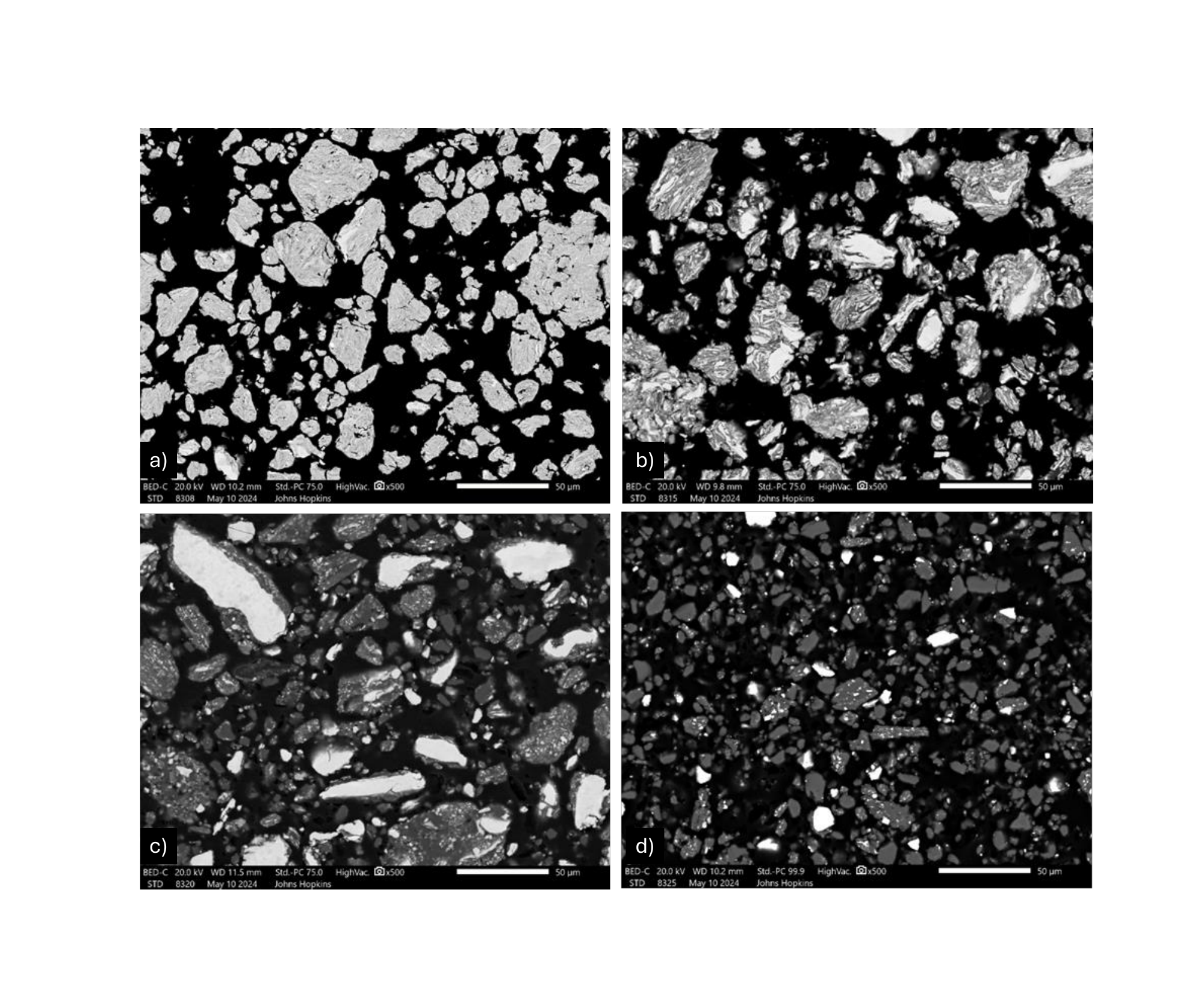}
    \caption{SEM cross sections for (a) (Al-8Mg):3Zr, (b) (Al-8Mg):Zr, (c) 2(Al-8Mg):Zr, and (d) 4(Al-8Mg):Zr powders.}
\label{Supplementary2}
\end{figure*}

\begin{figure*}[!htbp]
    \centering
    \includegraphics[trim=4.5cm 5cm 3.5cm 3.5cm, clip=true,width=.95\linewidth]{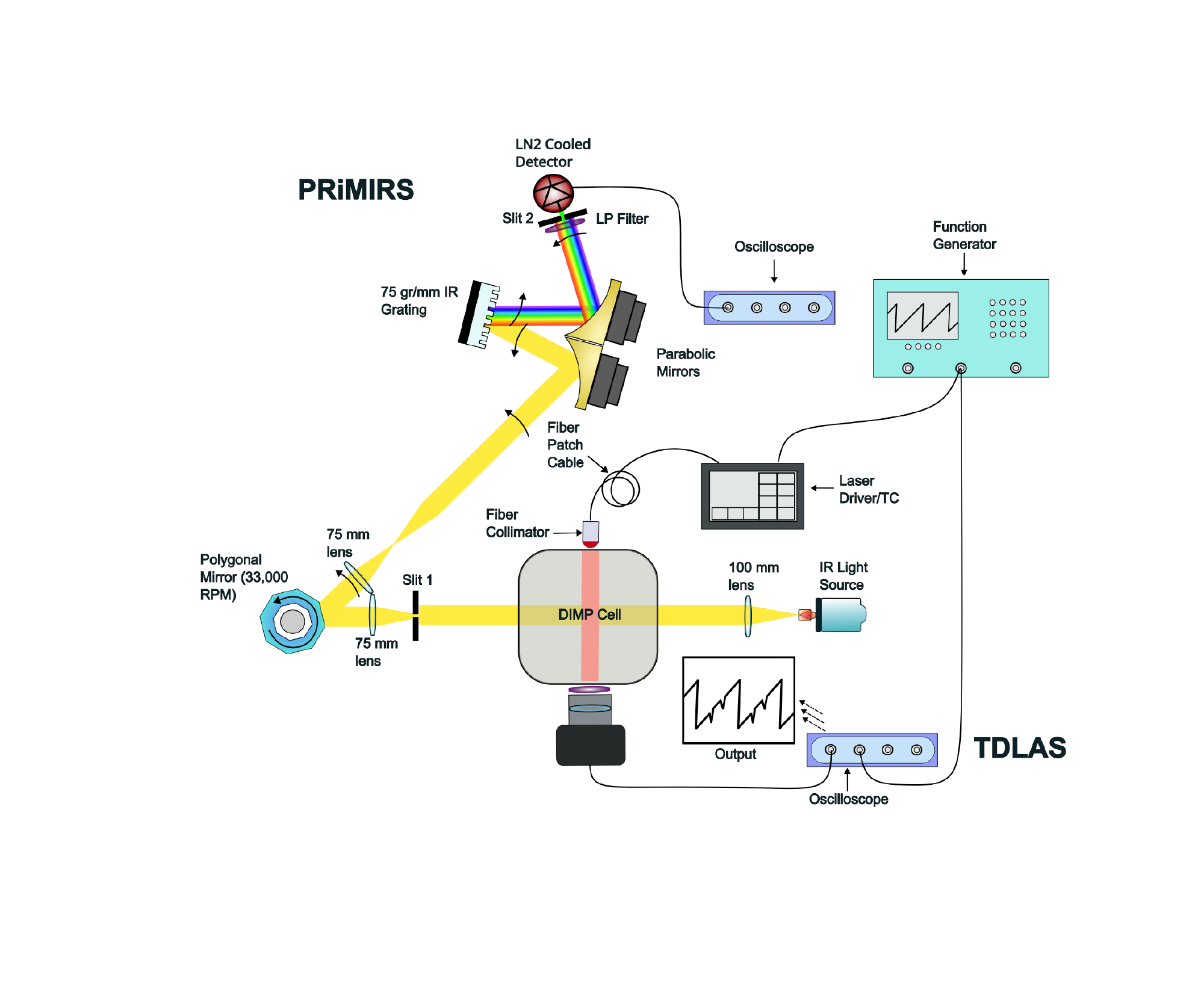}
    \caption{Schematic diagram of the Combined PRiMIRS and TDLAS spectrometers developed for simultaneous measurements of the IR spectral evolution of DIMP and background temperature profile due to combusting metal powders at 100 Hz or greater to measure variable defeat of CWAs.}
    \label{fig:PRIMIRSTDLAS}
\end{figure*}

A modular experimental chamber was constructed from a 6" x 6"x 6" aluminum frame (IdealVac P106861), with plates containing the following elements: Four KF-16 ports (top: P106866), Blank Plate (bottom: P108195), and four plates containing a port to load 1.0" standard optical windows of choice (sides: P108315). Ports on the top plate are outfitted with a KF16 to 1/8” NPT Swagelok(R) needle port to allow for sample loading using a syringe, a KF16 electrical conductor feedthrough to initiate the reaction (Kurt J Lesker EFT0021038Z) and blank KF16 flanges. The sides of the chamber include two 1" ZnSe wedged windows with anti-reflective coatings for 7-12 $\mu$m (Thorlabs WW71050-E3) aligned with PRiMIRS, and transverse to these are two 1" sapphire glass wedge windows (Thorlabs WW31050) aligned with TDLAS. The reactor is wrapped in heating tape (BriskHeat BS0051080L) and insulated using polyurethane foam sheets (Mcmaster 9385K61) to maintain thermal equilibrium. A chamber schematic of the observation ports, DIMP loading zone, powder loading, and electrode initiators is shown in \cref{fig:DIMPreactor}.

Composite metal powders prepared using a Retsch planetary mill were used for the experiments performed in this study. More details on this technique can be found in Vummidi \textit{et al}. \cite{VummidiLakshman2019ThePowders}. Ratios of an atomic percent (Al-8\%Mg) alloy from Goodfellow and Zr starting powders from Atlantic Equipment Engineers were milled for 45 minutes at 400 rpm with a ball to powder mass ratio of 3 using 10 mL hexane as the process control agent, then sieved below 75 $\mu$m. The specific formulations made for this study were: (Al-8Mg):3Zr, (Al-8Mg):Zr, 2(Al-8Mg):Zr, and 4(Al-8Mg):Zr and SEM cross sections are shown in \cref{Supplementary2}. These chemistries were chosen to vary the amount of material that will burn in the condensed phase (Zr) versus the amount of material that will burn in the vapor phase (Al-8Mg). Metal powders were weighed and spread across prepared 3Al-2Ni nanolayered foils of approximately 50 $\mu$m in total thickness whose fabrication process can be seen in Trenkle \textit{et al}.\cite{Trenkle2010Time-resolvedFoils} The reactive foil rests on a stainless steel block and push-pin electrodes, which connect to a power supply through a KF16 electrode feedthrough. Once the reactor is loaded with the designated powders, the cell is preheated to 90 °C and held for $\geq$10 minutes for thermal gradients to equilibrate within the cell. Background measurements are taken in order to obtain absorption spectra using PRiMIRS. 200 $\mu$l of liquid DIMP is then loaded through a syringe and allowed to generate DIMP vapor.

When the DIMP vapor reaches a relative signal maximum (>3.5 min), a Matlab program begins data acquisition and triggers the experimental sequence for the experimental setup shown in \cref{fig:PRIMIRSTDLAS}. Two oscilloscopes are triggered to simultaneously capture live data from both PRiMIRS (IR absorption by DIMP and its decomposition products) and TDLAS (Temperature) spectrometers at 100 Hz. After a brief delay (1 second), a current is sent through push-pin electrodes to initiate the reactive multilayer 3Al-2Ni foil. The foil provides the requisite heat to ignite metal powders, which then combust in the chamber containing DIMP vapor. Data is collected continuously for the full duration of prompt defeat (5s). Picoscope 6 waveform software is used to record spectra collected from each oscilloscope (Picoscope 5443D), one collecting from PRiMIRS and the other from TDLAS. The acquisition cycle is repeated with each turn of the polygonal mirror used in PRiMIRS to generate a new spectrum. For data acquisition within the Picoscope software, two particular measurement parameters were set to capture the desired information at the required time scales for both PRiMIRS and TDLAS. More details describing setting selection can be seen in Borah et al. \cite{Borah2024DevelopmentEnvironment}. The specific software parameters selected for oscilloscopes measuring PRiMIRS and TDLAS are shown in \cref{PicoscopeParameters}:

\subsection{Fourier Transform Infrared Spectroscopy}

For select experiments, the DIMP reactor was configured with a Bruker Invenio S FTIR with TGA-IR module as shown in \cref{fig:FTIRsetup}. A stainless steel transfer line was assembled using 1/4" swagelok fittings and tubes connecting the reactor to the TGA - FTIR unit comprising a stainless-steel gas cell with a temperature control unit. The IR beam reaches the cell through layered KBr/ZnSe windows. The beam path in the spectrometer housing outside the gas cell is flushed with dry air (by Airgas) with a flow rate of 3.5 L/min to minimize the effect of fluctuating humidity in the ambient air. A DLaTGS detector is used to analyze the gases, and absorption spectra are recorded in the range of 400 to 4500 cm$^-$$^1$. The transfer line and the TGA - IR module were preheated to 200°C to prevent condensation of DIMP and its decomposition products during transfer from the reaction vessel and during analysis. The sampling gas flow is maintained using a vacuum pump (JB Industries, model 0808) downstream of the TGA-IR module. The flow rate through the TGA-IR module is set to 0.2 L/min, using a needle valve and measured by an Alicat M series mass flow meter.

The reactor was loaded with (Al-8Mg):Zr powder placed on the 3Al-2Ni foil, ensuring good contact between the push-pin electrodes and the foil, and then reactor was preheated to 90 °C. Once stabilized, 200 $\mu$l of liquid DIMP was loaded into the experimental cell. Prior control experiments without igniting metal powder demonstrated that approximately 10 minutes was the appropriate residence time for liquid DIMP to volatize reaching a stable signal in the FTIR, so this was chosen as the time of ignition after loading DIMP. Due to difficulty in gas mixing within the chamber, swagelok needle valves needed to be opened approximately 30 seconds after ignition to allow the vacuum pump attached to the FTIR gas cell to adequately sample gases from within the DIMP reactor. FTIR measurements were made at 1.4 cm$^-$$^1$ resolution and continuously collected over 30 minutes with 20 seconds intervals post-reaction for verification of post-reactant products outside the range of PRiMIRS. Multiple experiments were performed using powders loadings that ranged from 125 to 200 mg.

\subsection{Transmission Electron Microscopy}

Gold TEM grids with a lacey-carbon film were placed throughout the DIMP reactor as shown in \cref{fig:TEMsetup}. 125 mg of (Al-8Mg):Zr powder was loaded onto 3Al-2Ni foil and ignited. DIMP was not loaded to protect the TEM from organic material. After initiating the reaction, the grids were collected and observed using TEM and STEM-EDS. A JEOL 200F 200kV TEM was used to obtain the bright field images of the metal oxides combustion products and observe their dispersion in the chamber. STEM-EDS was used to obtain the compositional mapping of the combusted particles.

\section{Results}

\subsection{Prompt-Defeat of DIMP evaluated using simultaneous operation of PRiMIRS and TDLAS systems}

\begin{figure*}[!htbp]
    \centering
    \includegraphics[trim=5.0cm 4.0cm 5.0cm 4.0cm, clip=true,width=0.97\linewidth]{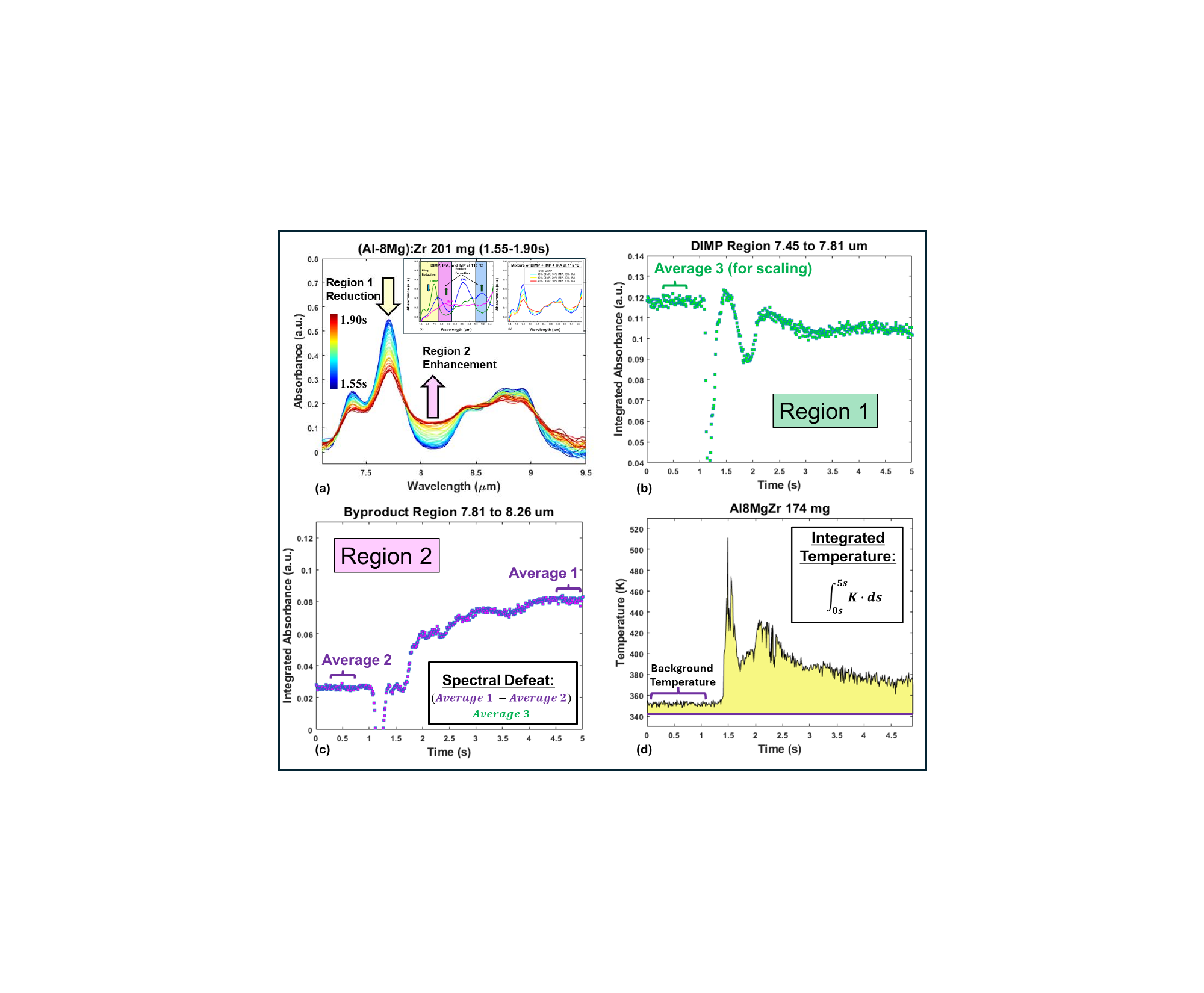}
    \caption{(a) Fast transformation of DIMP spectra captured by PRiMIRS demonstrating the need for 100 Hz time resolution. A small subset of expected decomposition product signal is shown adapted from \cite{Borah2024DevelopmentEnvironment}  (b) Region 1 (DIMP) integrated values are shown. (c) Region 2 (byproducts) integrated values are shown along with an equation for the DIMP spectral defeat parameter. (d) Temperature profile of global gas temperature calculated from path-averaged transmission spectra using TDLAS. Method for calculating the thermal contribution using integrated temperature is shown.}
    \label{fig:Dataprocessing}
\end{figure*}

\begin{figure*}[!htbp]
    \centering
    \includegraphics[trim=3.2cm 0cm 3.5cm 0cm, clip=true,width=1\linewidth]{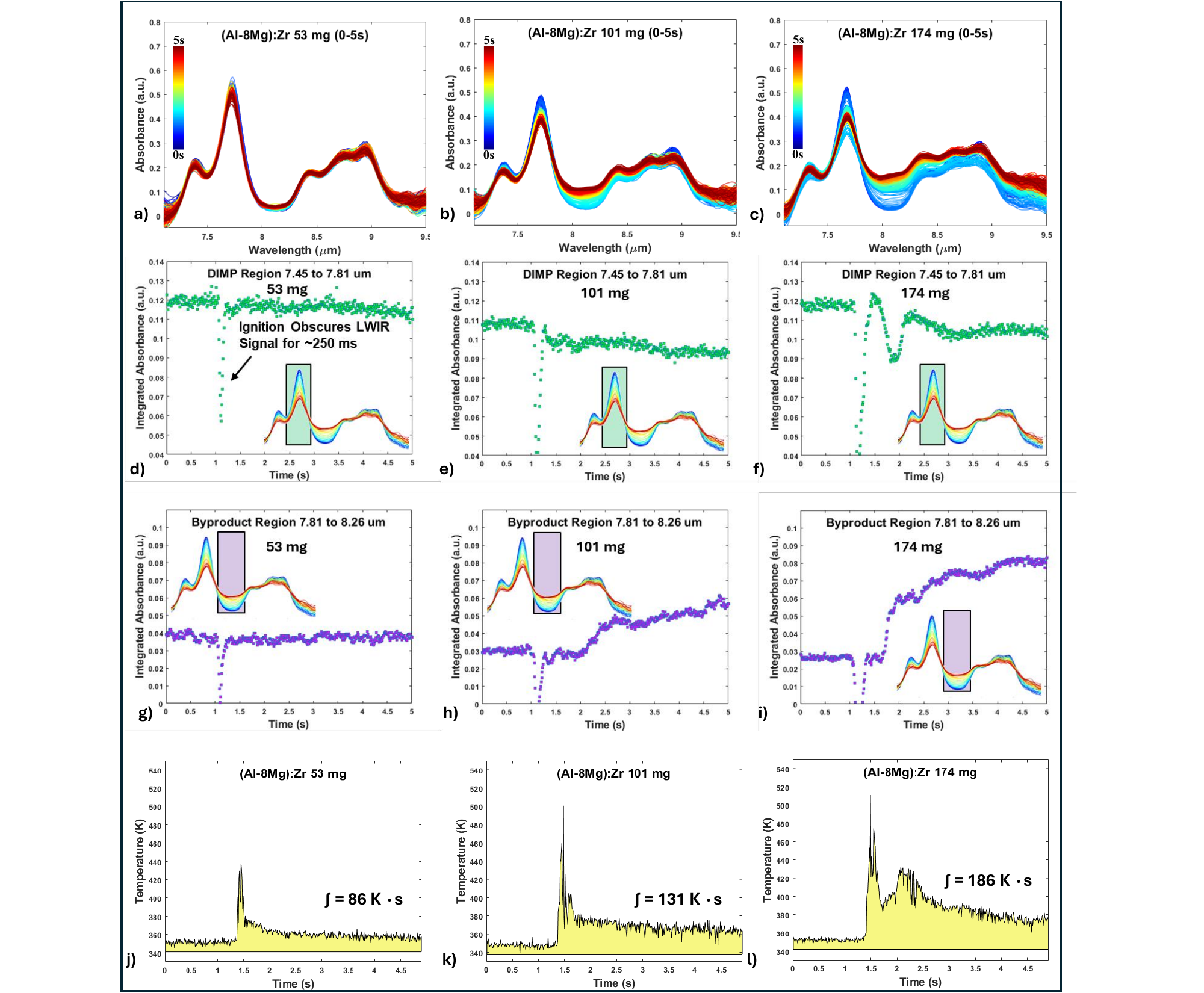}
    \caption{(a-c) Evolution of DIMP spectra captured at 100 Hz while igniting increasing loading amounts of (Al-8Mg):Zr powder where each plot is unique to a trial with a specific mass loading. (d-f) Integrated absorbance spectra for Region 1 from Fig. 7(a-c) for which the strongest contribution is attributed to DIMP. (g-i) Integrated absorbance spectra for Region 2 from Fig. 7(a-c) for which the strongest contribution is attributed to decomposition byproducts. (j-l) Temperature profiles measured during experiments shown in (a-c) via TDLAS method; the yellow shaded region corresponds to the integrated thermal parameter which is listed within the plots.}
    \label{fig:Maindata}
\end{figure*}

\begin{figure*}[!htbp]
    \centering
    \includegraphics[trim=1.0cm 7cm 0.5cm 7cm, clip=true,width=1\linewidth]{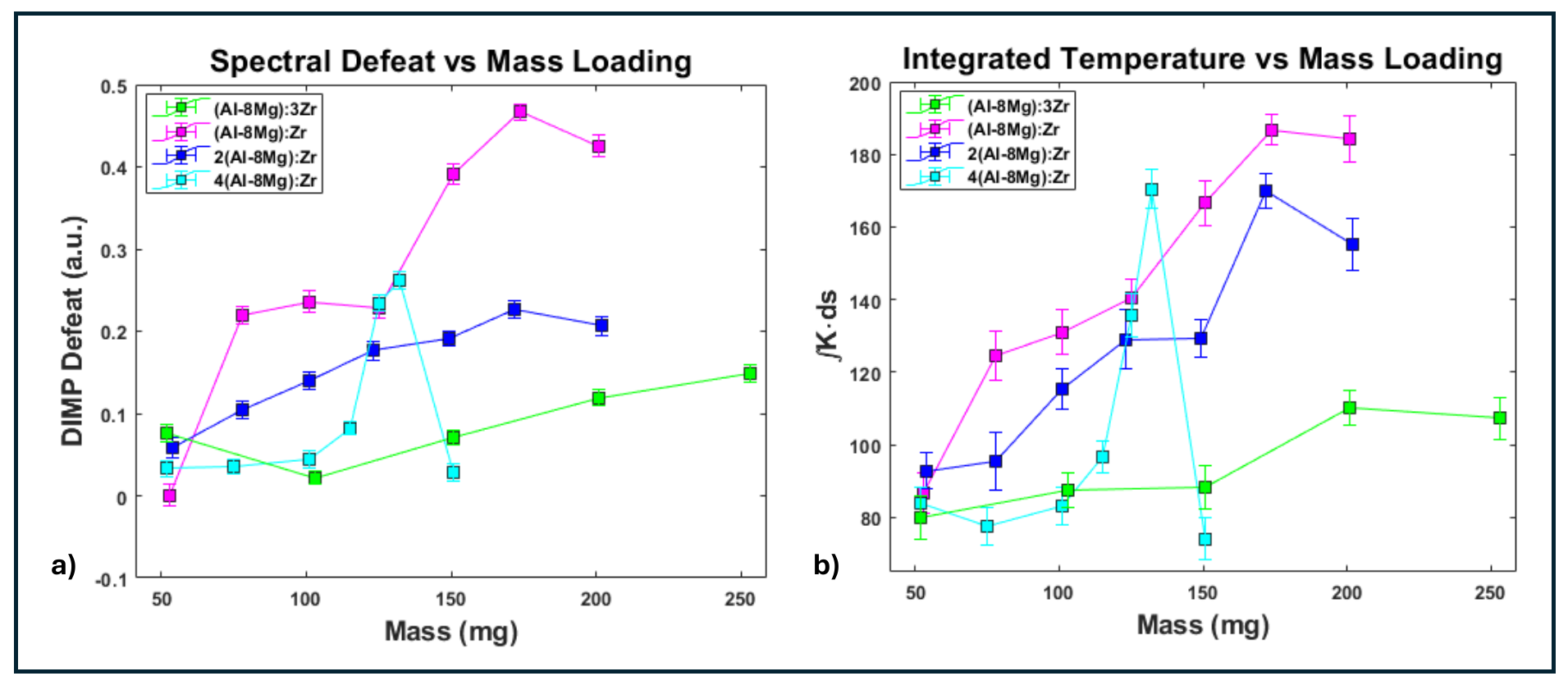}
    \caption{(a) DIMP spectral defeat vs. mass loading of different stoichiometric variations of (Al-8Mg)-Zr metal powder calculated from spectral data shown in \cref{fig:Dataprocessing}c. (b) Integrated temperature vs. mass loading of different stoichiometric variations of (Al-8Mg)-Zr metal powder calculated from spectral data shown in \cref{fig:Dataprocessing}d. Associated error is presented in \cref{Supplementary3}.}
    \label{fig:MassvsDefeat}
\end{figure*}

\begin{figure*}[!htbp]
    \centering
    \includegraphics[trim=5.5cm 7cm 5.5cm 7cm, clip=true,width=1\linewidth]{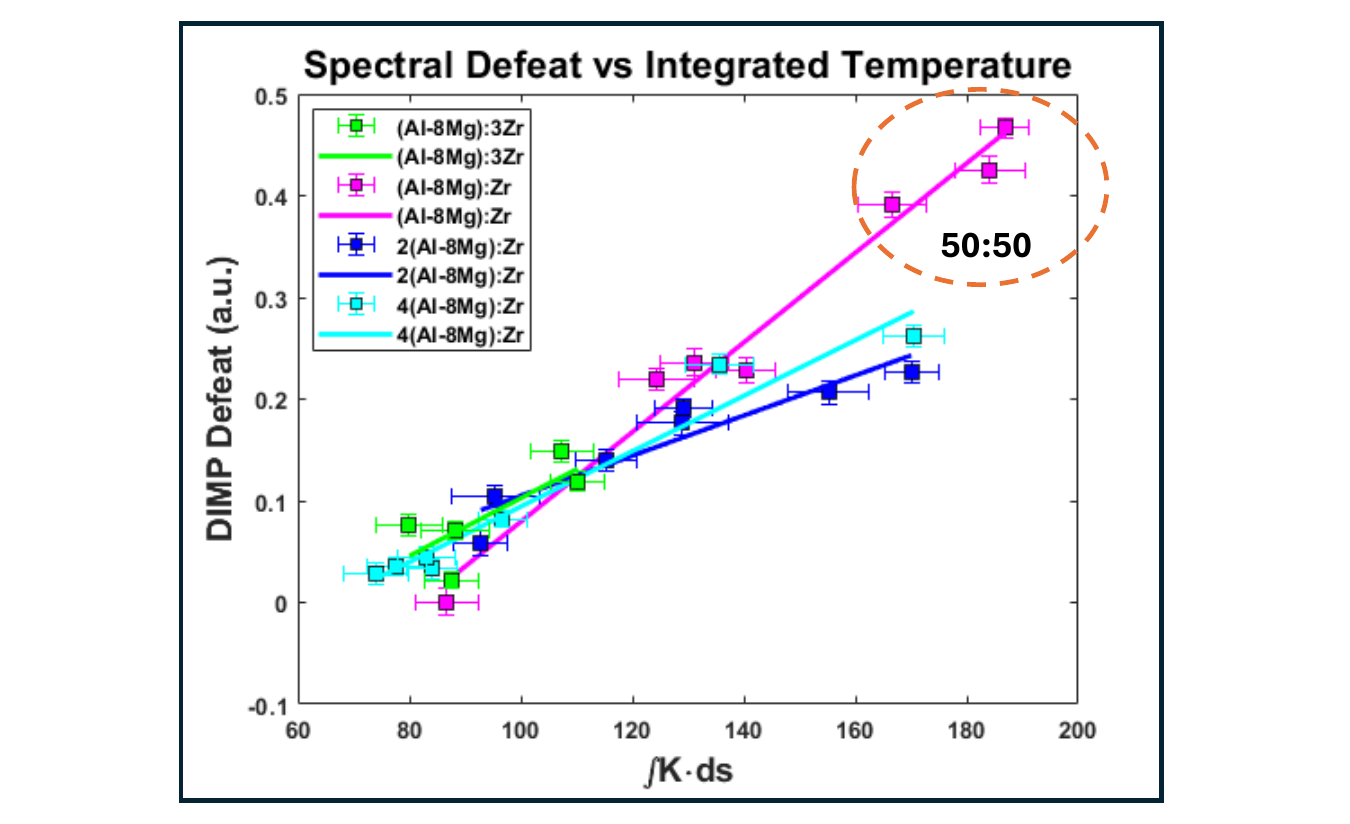}
    \caption{(a) DIMP spectral defeat vs. Integrated Temperature calculated from TDLAS measurements for stoichiometric variations of (Al-8Mg)-Zr powders. Respective slopes are with R-square values are available in \cref{Supplementary4}.}
    \label{fig:TemperaturevsDefeat}
\end{figure*}

A quantification method was previously proposed in Borah \textit{et al}. \cite{Borah2024DevelopmentEnvironment} identifying two key long-wave infrared (LWIR) absorption regions relevant to the decomposition of DIMP within the spectral range of PRiMIRS. Region 1 (7.5 to 7.8 $\mu$m) corresponds to DIMP, while Region 2 (7.8 to 8.3 $\mu$m) is attributed to products. Seen in \cref{fig:Dataprocessing}a and previously described in Borah et al. \cite{Borah2024DevelopmentEnvironment} where signal of products such as IMP and IPA appear in Region 2. PRiMIRS effectively captures the rapid evolution of the spectrum occurring during combustion events, as shown in \cref{fig:Dataprocessing}a demonstrating its ability to observe changes in DIMP and product spectral signatures at a resolution of 100 Hz. The 100 Hz spectra are generated through spectral averaging from measurements acquired at 1.3 kHz for SNR reduction. Additionally, \cref{fig:Dataprocessing}d shows the global gas temperature profile obtained from a DIMP + Metal Powder Combustion experiment extracted from path-averaged absorption spectra operated at 100 Hz using TDLAS. Integrated absorbance values for Regions 1 and 2 are shown in \cref{fig:Dataprocessing}b and \cref{fig:Dataprocessing}c, respectively. Because these regions contain overlapping IR signal from multiple products, concentration is difficult to extract. Therefore, in order to quantify different levels of decompostion and evaluate performance of different RMs, we introduce a comparative metric for decomposition known as the DIMP spectral defeat based on these measurements. The spectral defeat parameter is calculated by scaling the increase in the integrated absorbance of Region 2 (byproducts) to the initial integrated absorbance of Region 1 (DIMP). The final value for the integrated absorbance of Region 2 is calculated by averaging 50 data points at the end of the 5-second measurement, and the initial values are determined by averaging 50 data points recorded before the ignition of metal particles for Region 1 and Region 2. The thermal parameter, referred to as Integrated Temperature, was evaluated using trapezoidal integration of processed temperature data obtained via TDLAS.

\cref{fig:Maindata}a-c illustrates the spectral evolution of DIMP during the prompt defeat time regime (<5 s) while increasing (Al-8Mg):Zr loading amounts, measured using PRiMIRS at 100 Hz. As was described previously in Borah et al.\cite{Borah2024DevelopmentEnvironment} there is a brief period of $\sim$250 ms where our signal goes blind due to the combustion of metal powders releasing broadband IR light. Early time points are presented in blue, while later time points are presented in red. Integrated absorbances for Regions 1 (DIMP) and Region 2 (Byproducts), along with simultaneously measured temperature profiles for each experiment are shown in \cref{fig:Maindata}d-f, g-i, and j-l, respectively. Using these defined parameters, we report on the effect of mass loading between 50 and 250 mg on DIMP SDP in \cref{fig:MassvsDefeat}a, and Integrated temperature in \cref{fig:MassvsDefeat}b for metal powder chemistries: (Al-8Mg):3Zr, (Al-8Mg):Zr, 2(Al-8Mg):Zr, and 4(Al-8Mg):Zr. Increases in background temperature and DIMP spectral defeat with greater mass loadings were generally observed for all formulations tested within this study. Drops to both defeat and temperature are evident at higher mass loadings particularly in Al-rich formulations which will be elaborated on in the discussion section. To further analyze spectral defeat behavior while acknowledging variations in ignition and combustion efficiencies among powder chemistries, \cref{fig:TemperaturevsDefeat} plots Spectral Defeat vs. Integrated Temperature for each composition. Here we observe a strong trend between integrated temperature and increased DIMP spectral defeat. Slope evaluation using a first-order polynomial fit was used to quantify metal powder performance with spectral defeat evaluated as function of thermal output.  The calculated slopes for each chemistry are as follows: (Al-8Mg):Zr - 4.406e$^-$$^3$, 2(Al-8Mg):Zr - 1.972e$^-$$^3$, 4(Al-8Mg):Zr - 2.725e$^-$$^3$, and (Al-8Mg):3Zr - 2.841e$^-$$^3$ where the dimensions compare DIMP spectral defeat to integrated temperature and are presented in \cref{Supplementary4}.   \\

\subsection{FTIR verification of DIMP Decomposition Products outside the spectral range of PRiMIRS}

\begin{figure*}[!htbp]
    \centering
    \includegraphics[trim=2.5cm 0.0cm 2.5cm 0.0cm, clip=true,width=1\linewidth]{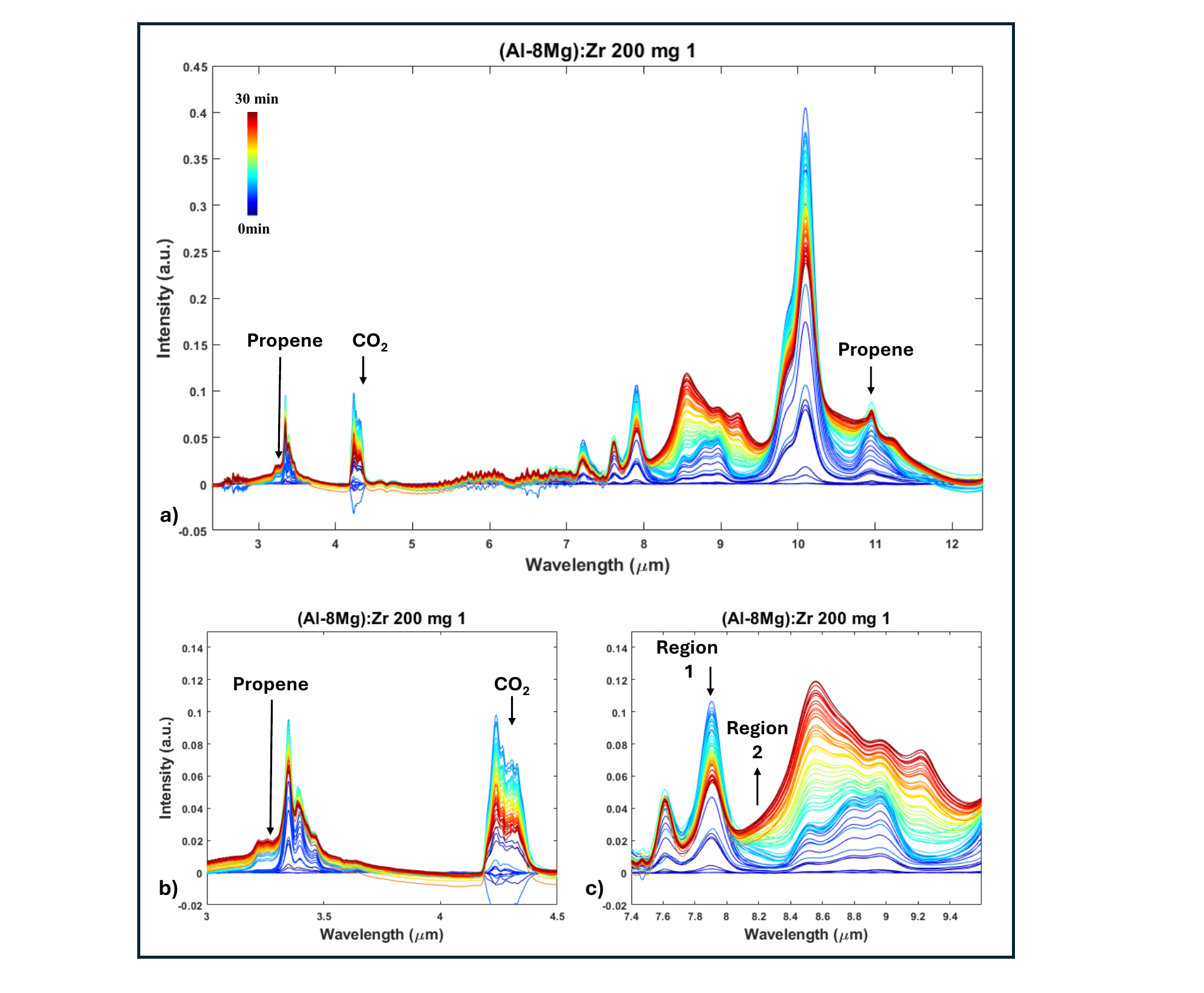}
    \caption{(a) FTIR spectra captured over 30 minutes from the time when DIMP is introduced into the heated chamber (darkest blue) and stabilizes (dark blue) through the ignition of (Al-8Mg):Zr powders and onto the formation of byproducts at later times (dark red). (b) FTIR signal evolution of expected byproducts outside the spectral range of PRiMIRS such as Propene (left) and CO$_2$ (right) are shown. (c) FTIR signal of the spectral range measured by PRiMIRS is shown. Here we also observe a reduction in signal within Region 1 (DIMP) and a rise in signal in Region 2 (byproducts) which correlate to byproduct signals shown in (b).}
    \label{fig:FTIRData}
\end{figure*}

While PRiMIRS excels at capturing rapid transitions relevant to combustion event timescales, its current configuration is less suited for a detailed analysis of specific decomposition products, which require a broader wavelength range. To complement PRiMIRS, FTIR was used to enhance confidence in the Spectral Defeat metric, calculated from the Region 1/Region 2 transition, by correlating PRiMIRS measurements with evidence of known decomposition products that exhibit spectral signatures outside its wavelength range. 

In \cref{fig:FTIRData}c we see a similar spectral evolution in the FTIR data as was observed with PRiMIRS. There is a reduction in signal in the region below 8 $\mu$m and an enhancement in signal within the region above 8 $\mu$m. Given that continous measurements were made over much longer time periods in the FTIR (30+ minutes) than in PRiMIRS, where measurements focused on prompt defeat timescales (<5s), observing a more significant spectral evolution was not surprising. \cref{fig:FTIRData}b highlights signal increases indicative of decomposition products propene with the rising shoulder below 3.3 $\mu$m and CO$_2$ at 4.2 to 4.4 $\mu$m. The full FTIR spectra is shown in \cref{fig:FTIRData}a. Here we have another indication of propene formation with the observed formation of a sharp peak at 11 $\mu$m. FTIR provided additional validation and increased confidence in the spectral defeat metric calculated from measurements made via PRiMIRS.

\subsection{TEM Verification of Expected Combustion Products from (Al-8Mg):Zr Powders}

\begin{figure*}[!htbp]
    \centering
    \includegraphics[trim=0cm 7cm 0cm 6.5cm, clip=true,width=0.95\linewidth]{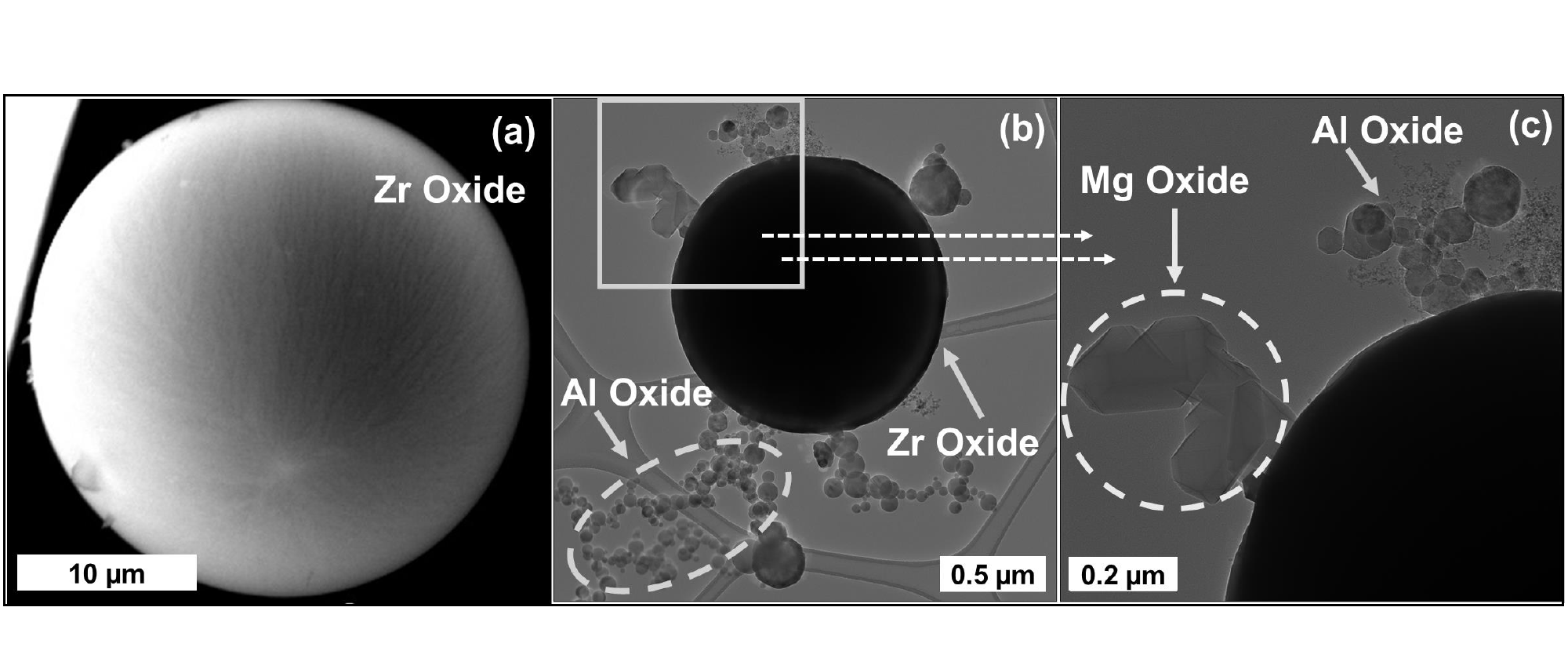}
    \caption{(a) SEM image of a large ZrO$_2$ particle. Due to the resolution limit of SEM, MgO and Al$_2$O$_3$ were imaged using TEM. (b) Bright field TEM image of all combustion products at lower magnification. (c) Magnified view of Al$_2$O$_3$ and MgO exhibiting spherical nature and sharper edges, respectively.}
\label{fig:TEMpicture}
\end{figure*}
\begin{figure*}[!htbp]
    \centering
    \includegraphics[trim=0cm 5cm 0cm 5cm, clip=true,width=0.95\linewidth]{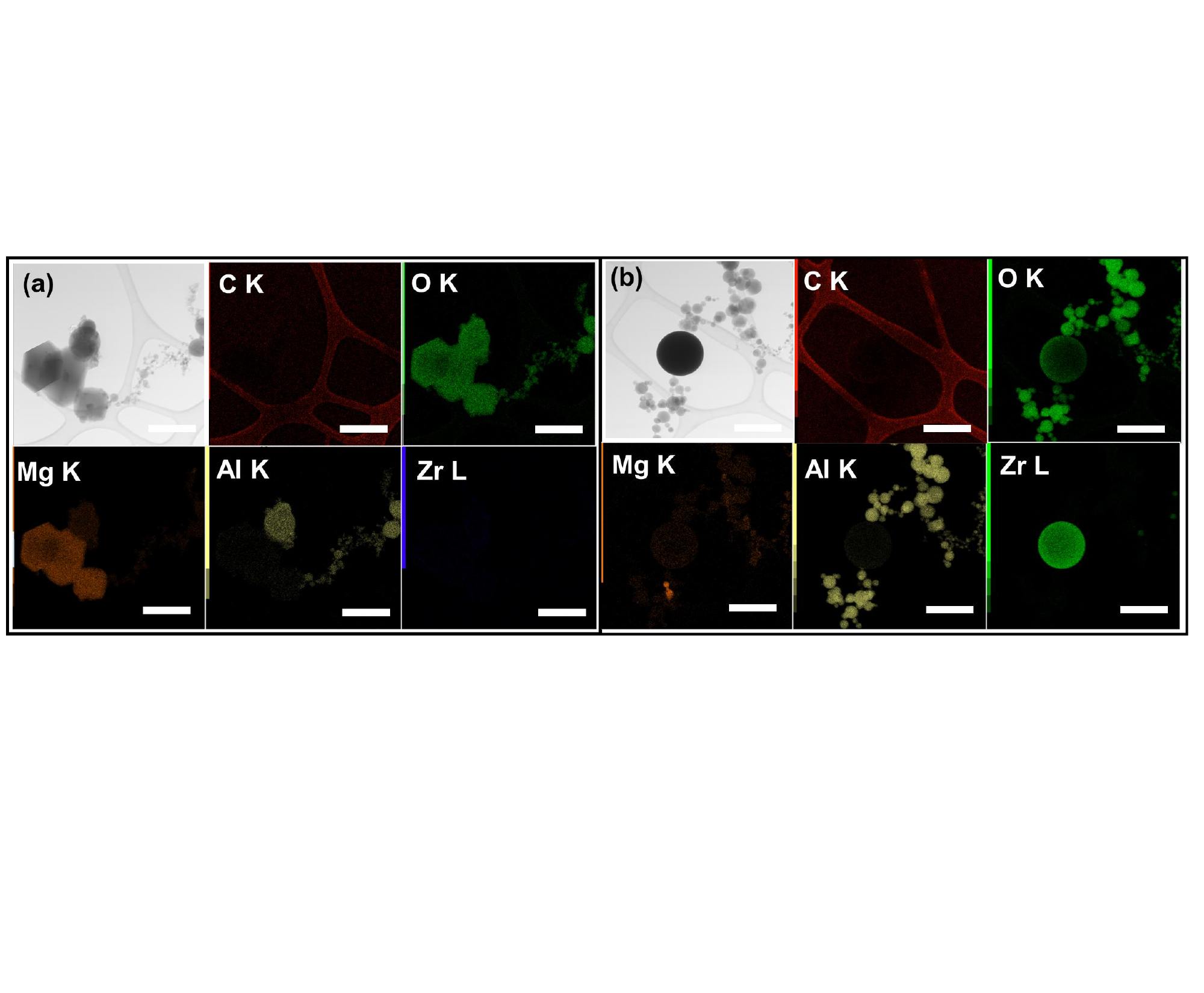}
    \caption{STEM-EDS elemental mapping of combustion products after igniting (Al-8Mg):Zr metal powder that were captured on a Au, Lacey Carbon TEM grid showing regions of (a) MgO and Al$_2$O$_3$ and (b) ZrO$_2$ and Al$_2$O$_3$ particles. Elemental information matches ZrO$_2$ to the larger dark spheres, Al$_2$O$_3$ to the smaller spherical particles, and MgO having sharper edges. The scale bar in each picture corresponds to 500 nm.}
\label{fig:EDSpicture}
\end{figure*}

The objective of this study was to investigate the variable decomposition of DIMP, with a particular attention to the influence of metal powder chemistry. To ensure that the combustion products from the metal powders actively contributed to the reaction environment containing DIMP vapor, it was crucial to verify their presence throughout the chamber. TEM grids were placed throughout the chamber as discussed in the Materials and Methods section and (Al-8Mg):Zr powders were ignited. We were looking for evidence of Al, Mg, and Zr in their oxidized form in different locations within the reaction space. Each grid contained evidence of metal oxides: ZrO$_2$, Al$_2$O$_3$, and MgO. These oxides varied in morphology and size as can be observed in \cref{fig:TEMpicture}. The ZrO$_2$ particles are large, dark spheres, the Al$_2$O$_3$ particles are small, light spheres, and the MgO particles are cubic with sharp edges. The diameter of ZrO$_2$ ranges from 500 nm to 30 um, while the Al$_2$O$_3$ are sub-100 nm in diameter. The elemental chemistries are further verified using EDS and shown in \cref{fig:EDSpicture}. This confirmation provided evidence that the reacted particles were present and traversed the reaction space.

\clearpage

\section{Discussion}

\subsection{TEM + EDS Confirmation of Metal Oxide Combustion Products}

The TEM images in \cref{fig:TEMpicture} and the STEM-EDS scans in \cref{fig:EDSpicture} demonstrate that each element in the (Al-8Mg):Zr composite powder formulation underwent combustion and formed oxide particles following ignition by the 3Al-2Ni foil. Observations from multiple grids dispersed throughout chamber confirm this conclusion and suggest the oxide products were dispersed throughout the chamber. As expected \cite{Wainwright2018ObservationsEnvironmentsb,Overdeep2015UsingNanolaminates}, the Zr oxide particles that are known to combust in the condensed phases are larger than the Al and Mg oxide particles that nucleate and grow in the vapor phase.  Mixed oxides were not observed on the collected grids, suggesting the majority of Al and Mg boil off from each combustion composite particle, yielding large, single metal zirconia oxide particles. The spherical nature of the Zr particles suggest they oxidize in the liquid state as expected \cite{Wainwright2018ObservationsEnvironmentsb}. The separate oxides for Al and Mg are attributed to their very different vaporization temperatures and the fact that they nucleate and grow in the vapor state. The much finer size of the Al and Mg oxide particles compared to the Zr oxide particles implies they yield far more surface area with which the DIMP can react during a neutralization experiment. We do note that the evaluated material represents a only fraction of the overall experiment, but the main objective of this experiment was to verify that RMs used during this experiment formed the expected combustion products previously observed in \cite{Wainwright2018ObservationsEnvironmentsb,Overdeep2015UsingNanolaminates} and were able to be captured further into the chamber space and off of the ignition block.\\

\subsection{Prompt Defeat of DIMP using Combined PRiMIRS  and TDLAS Spectrometers}

Increases in background temperature and DIMP defeat were generally observed for all formulations tested within this study. As shown in \cref{fig:MassvsDefeat}, we observe a general increase in both DIMP defeat and integrated temperature with greater amounts of mass loading across all chemistries. This trend is expected, as burning reactive materials generates energy, leading to greater thermal contributions and more DIMP defeat with increased mass loading. However, for higher loadings in some chemistries, we note a decline in both spectral defeat and integrated temperature, particularly in the aluminum-rich chemistries 2(Al-8Mg):Zr, and even more drastically in 4(Al-8Mg):Zr. This decrease can be attributed to the difficulty in igniting Al-rich formulations  \cite{VummidiLakshman2019ThePowders}, particularly at higher mass loadings, resulting in reduced combustion efficiency within the given experimental conditions. To maintain consistency, 3Al-2Ni foils of comparable mass, size, and thickness were prepared for each mass loading that varied from 50 to 250 mg for the four different chemistries. However, aluminum-rich formulations of equivalent mass exhibited a higher powder volume due to aluminum’s significantly lower density compared to zirconium. Consequently, this led to thicker metal powder beds in aluminum-rich formulations. The increased bed thickness may have limited the ability of the reacting 3Al-2Ni foils to effectively ignite the full Al-rich powder beds, in contrast to the more Zr-rich formulations where more complete ignition was observed. \\

\cref{Supplementary1} presents images of post-fire ignition blocks for various chemistries as loading amounts increase. As the formulation transitions from (Al-8Mg):Zr to 4(Al-8Mg):Zr, we observe evidence of less reaction and spreading of powders off the test blocks. For the 4(Al-8Mg):Zr powders in particular, a significant portion of the powder remains unreacted at higher loadings, retaining its original light gray coloration. Since Al-rich formulations primarily combust in the gas phase, we hypothesize that the initial gas production from burning material in direct contact with the multilayer foil pushes unreacted powder within the thicker powder bed away from the ignition source before sufficient heat can propagate to sustain ignition and subsequently combustion. In addition to loading amount, difficulty in ignition can also be due to chemistry and is understood by observing SEM cross sections of these composite reactive powders in \cref{Supplementary2}. In backscatter scanning electron microscopy, heavier elements appear brighter, while lighter elements appear darker—thus, Zr inclusions are bright, whereas the Al-8Mg matrix appears dark. \cref{Supplementary2}b, which corresponds to the 50:50 (Al-8Mg):Zr chemistry—the highest-performing formulation in DIMP defeat—reveals well-distributed Zr inclusions throughout individual particles, forming multiple bands interspersed within the Al-rich alloy. However for the (Al-8Mg)-rich samples in \cref{Supplementary2}c and especially \cref{Supplementary2}d, this uniform distribution is noticeably absent. Instead, particles tend to be either predominantly Al-rich or Zr-rich, rather than exhibiting a homogeneous dispersion of Zr content within each particle. This poor integration can hinder ignition propagation and combustion efficiency, as the intermetallic reaction between Al and Zr, which occurs during melting, plays a critical role in reducing ignition thresholds by generating localized heat. \cite{VummidiLakshman2019ThePowders,Fritz2013ThresholdsEnergy,Overdeep2015UsingNanolaminates,Wainwright2018ObservationsEnvironments} As a result, for Al-rich formulations (greater than 50:50), we observe significantly lower integrated temperatures and reduced DIMP defeat at higher loadings, as shown in \cref{fig:MassvsDefeat}. The decreased ignition efficiency of these formulations, even when using the same reaction initiator, results in a weaker overall combustion event and thermal pulse.               \\

Given the observed variations in ignition and combustion efficiencies across different stoichiometric formulations, we examined the relationship between integrated temperature and DIMP defeat. The premise was that by monitoring temperature, we could account for the relative strength of combustion events, thereby mitigating run-to-run and formulation-dependent variations. This approach allowed us to evaluate performance across different chemistries while acknowledging that their reaction behaviors differed. \cref{fig:TemperaturevsDefeat} reveals a strong correlation between DIMP defeat and integrated temperature, with (Al-8Mg):Zr emerging as a notably high-performing formulation. A key takeaway from this comparison is that, despite the ignition challenges associated with Al-rich chemistries, we can still assess their defeat potential in relation to their measured thermal contributions, or thermal efficiency, by evaluating their slopes. It is important to acknowledge that in Al-rich formulations there are limited data points at the higher temperatures achieved by (Al-8Mg):Zr, and the Zr-rich formulation remained at lower temperatures throughout. Thus, total spectral defeat is still important to consider despite differences in the calculated slopes. The greatest slope was exhibited by the 50:50 (Al-8Mg):Zr chemistry, followed by (Al-8Mg):3Zr, 4(Al-8Mg):Zr, and lastly 2(Al-8Mg):Zr. A regression model was run utilizing Matlab to obtain Confidence intervals for the calculated slopes for each chemistry and are shown in \cref{Supplementary4}. The Zr-rich formulation has the lowest R-square value in slope due to a low distribution in observed defeat. This accounts for the unusually large confidence interval compared to the other chemistries and thus we will not comment or compare this slope with the remaining formulations. The confidence intervals for the slopes in 2(Al-8Mg):Zr and 4(Al-8Mg):Zr do overlap so despite the higher observed slope in 4(Al-8Mg):Zr we cannot strongly consider it to be a better performer. However, the confidence intervals for both of these chemistries fall outside the confidence interval of (Al-8Mg):Zr. This supports our determination of (Al-8Mg):Zr to be the best performer in this study having demonstrated the best overall magnitude in spectral defeat and thermal efficiency determined by slope. \\

The primary objective of exploring stoichiometric variation was to investigate DIMP defeat by increasing particle interactions—transitioning from condensed-phase generators (Zr-rich) to vapor-phase generators (Al-rich). However, due to the ignition difficulties discussed earlier, we cannot definitively conclude whether increased particle interactions were achieved as formulations became more Al-rich content. We qualitatively suspect a relatively complete reaction from the Zr-rich chemistry by observing images of post-ignition blocks shown in \cref{Supplementary1}. Because it predominantly burns in the condensed phase, the resulting solid oxide remains stationary post-fire and does effectively disperse away from the ignition block. In contrast, the 50:50 formulation exhibits a more radial distribution originating from the 3Al-2Ni foil initiator. Given the difficulties in ignition with 4(Al-8Mg):Zr and the limited number of experiments approaching moderate defeat levels ($\sim$0.15 DIMP defeat or greater) in both 4(Al-8Mg):Zr and (Al-8Mg):3Zr, our most reliable comparison lies between (Al-8Mg):Zr and 2(Al-8Mg):Zr where the 50:50 chemistry is the best performer in both total defeat and slope (thermal efficiency). \\

Our initial expectations were that by increasing surface interactions we could enhance DIMP defeat. This held true for our 50:50 chemistry (Al-8Mg):Zr against the Zr-rich (Al-8Mg):3Zr indicating that vapor-phase generation may play a crucial role in optimizing formulations for agent defeat applications. However, (Al-8Mg):Zr outperforming 2(Al-8Mg):Zr and 4(Al-8Mg):Zr is counter-intuitive to this hypothesis, as the more Al-rich formulations should produce even more vapor phase products. While temperature measurements were made in the same location, it is possible that dispersal beyond this location suffered or smoke generation varied between formulations in addition to ignition challenges as fine particulate residue on chamber surfaces was most evident using the 50:50 (Al-8Mg):Zr chemistry. This would suggest that surface interactions between combustion products and DIMP may not have been successfully increased in the Al-rich formulations during this experiment. To effectively compare different vapor-phase generators going forward, it is essential to establish a consistent and reliable ignition method across formulations, as well as to quantify combustion product dispersal and smoke generation.   

\subsection{FTIR Verification of DIMP Decomposition Products}

While PRiMIRS provides strong temporal resolution for assessing prompt defeat using LWIR spectroscopy, its inability to capture a broader wavelength range limits the observation of additional decomposition species. This raises questions about the spectral defeat parameter calculated using the Region 1/Region 2 transition shown in \cref{fig:Dataprocessing}. In \cref{fig:FTIRData}a we present continuous FTIR data collection of DIMP volatilizing within the preheated chamber. The DIMP signal continually increases until it reaches a peak, at which point a metal powder reaction is initiated using (Al-8Mg):Zr, followed by subsequent spectral measurements. The extended collection time compared to PRiMIRS was to address a noticeable delay between species formation in the reactor and their detection by FTIR. This delay suggests challenges in gas mixing within the DIMP reactor, as the heated transfer line is connected to the upper corner of the reaction vessel. In contrast, PRiMIRS and TDLAS employ optical paths that measure through the chamber’s center, near the reaction zone, where local transformations occur and can be readily observed. Since the FTIR system could not be positioned for in-situ measurements within the reactor, generated species must first equilibrate and migrate toward the heated transfer tube before they can be detected. As a result, we do not use FTIR data to comment on time-resolved species evolution. However, we can confirm that specific IR signatures corresponding to DIMP decomposition products—beyond the spectral range of PRiMIRS—are observed post-fire. \\

As highlighted in \cref{fig:FTIRData}b, we identify two key locations where spectral signals indicate the presence of explicit decomposition products—propene and CO$_2$. A distinct rise appears near 3.3 $\mu$m after the dark blue (pre-fire) spectra, corresponding to propene formation. \cite{Yuan2019T-jumpMethylphosphonate,Senyurt2024ExperimentalDIMP} Similarly, a sharp increase in CO$_2$ absorption at 4.3 $\mu$m is observed. \cite{Yuan2019T-jumpMethylphosphonate,Senyurt2024ExperimentalDIMP} However, because gas mixing was stimulated by opening valves to facilitate transport between the reactor and the FTIR, CO$_2$ -a highly volatile species- exhibits a declining signal over time, likely due to its easy escape under these conditions. Additionally, \cref{fig:FTIRData}a reveals another indication of propene formation, marked by a sharp peak at 11 $\mu$m \cite{Yuan2019T-jumpMethylphosphonate,Senyurt2024ExperimentalDIMP}. Turning to \cref{fig:FTIRData}c, we observe a spectral transformation within the PRiMIRS wavelength range, where the signal in Region 1 (DIMP) decreases while the signal in Region 2 (byproducts) increases. Since these measurements extend well beyond PRiMIRS’ 5-second capture, they demonstrate a greater degree of transformation, capturing changes that occur beyond the prompt defeat regime. The spectral shifts used to assess the prompt defeat of the CWA simulant DIMP via PRiMIRS are also evident in the FTIR data, with clear signals corresponding to decomposition products such as propene and CO$_2$. This correlation strengthens confidence that the spectral defeat observed in PRiMIRS is indeed indicative of DIMP decomposition, further validated by byproduct formation detected across a broader wavelength range.      \\

\section{Conclusion}

This study aimed to demonstrate a method for evaluating and quantifying the prompt defeat capabilities of reactive powders using the simultaneous operation of two spectrometers, PRiMIRS and TDLAS. We successfully observed prompt defeat of CWA simulant DIMP while combusting metal powders. Thermal pulse and DIMP defeat generally increase with mass loading, however variations in ignition and combustion performance led us to investigate DIMP defeat vs integrated temperature. This allows us to account for chemistries demonstrating poor ignition and/or combustion and still evaluate performance. While the defeat does rise with increased thermal pulse across all chemistries, the slope of this increase was greatest for (Al-8Mg):Zr. 

To explore this, we varied the stoichiometric ratios of an (Al-8Mg)-Zr-based formulation, hypothesizing that vapor-phase generators (Al-8Mg)-rich formulations would promote greater surface interactions (greater smoke), whereas condensed-phase Zr-rich generators would exhibit reduced surface interactions (less smoke). Our results demonstrate a strong thermal dependence on prompt defeat, where despite ignition and combustion challenges, an approximately linear relationship emerges between increased thermal behavior and enhanced defeat - an expected trend as enhanced combustion events provide more heat and combustion products (smoke) to the measured area. Among the tested formulations: (Al-8Mg):3Zr, (Al-8Mg):Zr, 2(Al-8Mg):Zr, and 4(Al-8Mg):Zr, the (Al-8Mg):Zr formulation stands out as the highest performer, both in magnitude of Spectral Defeat and slope indicative of thermal efficiency. Contrary to our hypothesis however, while performance increased going from Zr-rich to 50:50, this trend did not continue for Al-rich formulations beyond 50:50.

This work establishes a quantitative comparative method for evaluating prompt defeat performance across multiple formulations and presents the first measurements on variable prompt defeat using reactive materials in a non-high-explosive setting. Moving forward, we aim to further investigate the influence of chemical and smoke variation on CWA simulant defeat by selecting formulations with different elemental metal oxides rather than varying the stoichiometric ratios of the same elements.     \\

\section{Declaration of competing interest}
We, the authors, confirm we have no financial or personal conflicts that could have influenced the findings presented in this study.

\section{Data Availability Statement}
The data that support the findings of this study are available within the article and its supplementary material. Additional data are available from the corresponding authors upon reasonable request.

\section{Acknowledgments}

This work was supported by the Department of Defense, Defense Threat Reduction Agency (DTRA) under the MSEE URA, HDTRA1-20-2-0001.  

\vspace{10pt}

\clearpage
\appendix
\onecolumngrid
\section{Supplementary Information}

\setcounter{figure}{0}
\renewcommand{\figurename}{A}
\subsection{Oscilloscope Parameters}
\label{PicoscopeParameters}
\begin{enumerate}
    \item TDLAS - The capture parameter is a continuous collection of raw spectral measurements collected over 5 seconds where spectra are generated on a 50\% ramp that is repeated at 100 Hz. Within the Picoscope software, the number of samples is set to 5 MS (million) and the timebase is 500 ms/div.
    \item PRiMIRS - Operating at a capture rate of 1.3 kHz, 13 sequential spectra are averaged resulting in 100 Hz temporal resolution with signal to noise ratio (SNR) enhancement. Picoscope settings were adjusted to the following parameters: 500 waveforms were collected with rapid pulse technology ($\leq \mu$s delay between waveforms); each waveform contained 50,000 samples and the timebase was set to 1 ms/div.
\end{enumerate}

\subsection{FTIR Setup}
\label{fig:FTIRsetup}
\begin{figure*}[!htbp]
    \centering
    \includegraphics[trim=3.5cm 9.5cm 3.5cm 6.5cm, clip=true,width=1\linewidth]{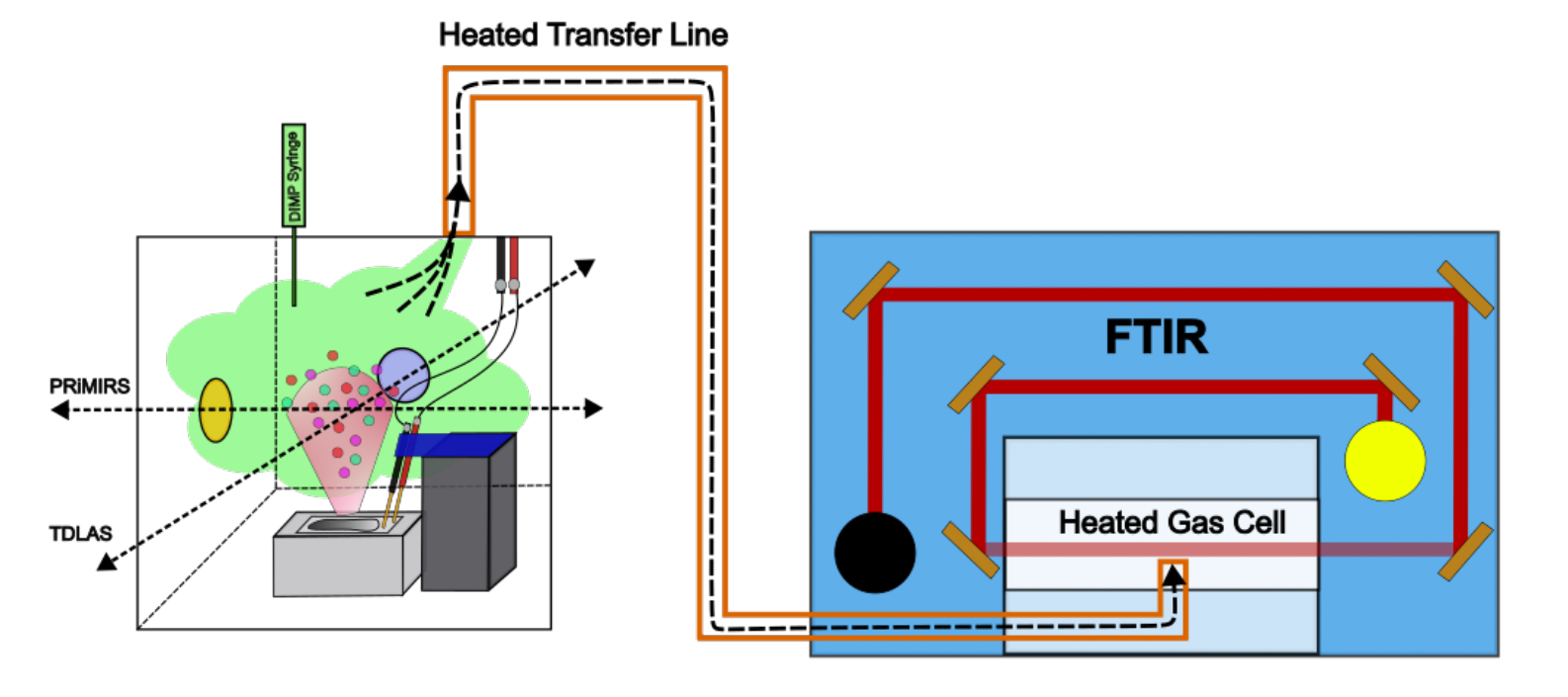}
    \caption{Experimental chamber shown in \cref{fig:DIMPreactor} is connected to a Bruker FTIR through a SS heated transfer line.} 
\end{figure*}
\clearpage
\subsection{TEM Setup}
\label{fig:TEMsetup}
\begin{figure}[h]
    \centering
    \includegraphics[trim=8cm 7cm 8cm 7cm, clip=true,width=0.6\linewidth]{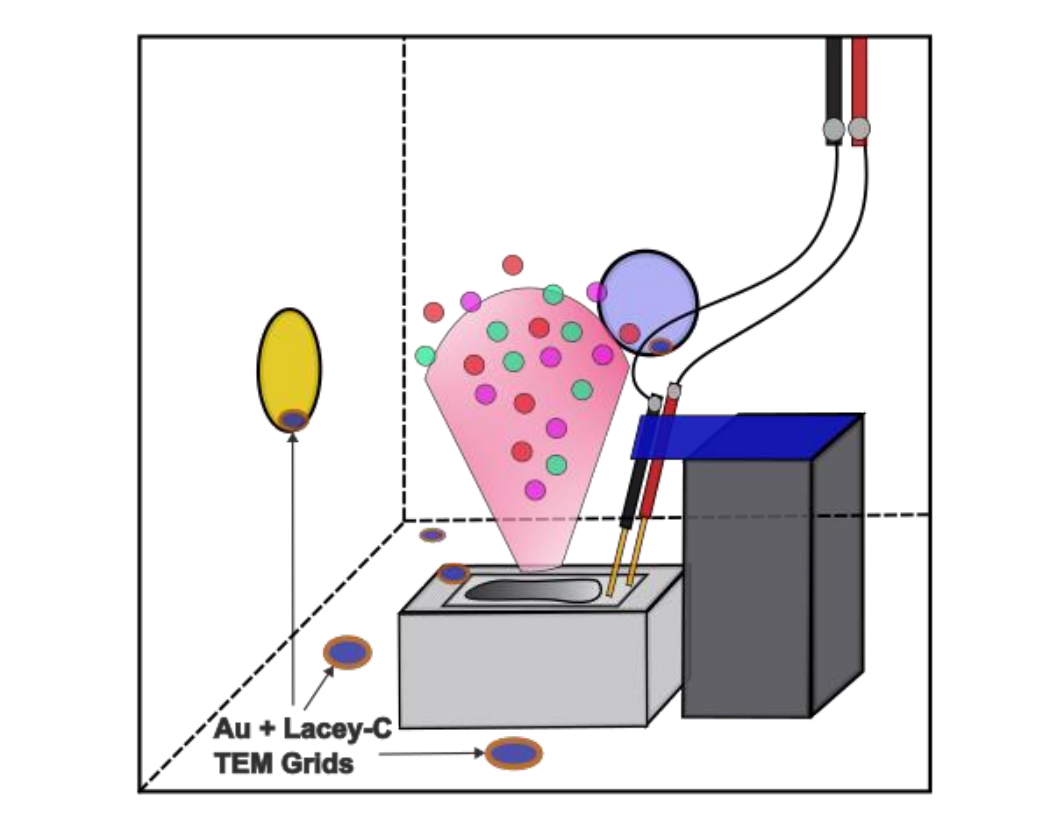}
    \caption{Schematic of TEM grid placement inside DIMP reactor}
\end{figure}

\subsection{Post-fire Ignition Blocks for Varying Chemistries and Loading Amounts}
\label{Supplementary1}
\begin{figure*}[h]
    \centering
    \includegraphics[trim=0cm 6cm 0cm 6cm, clip=true,width=0.95\linewidth]{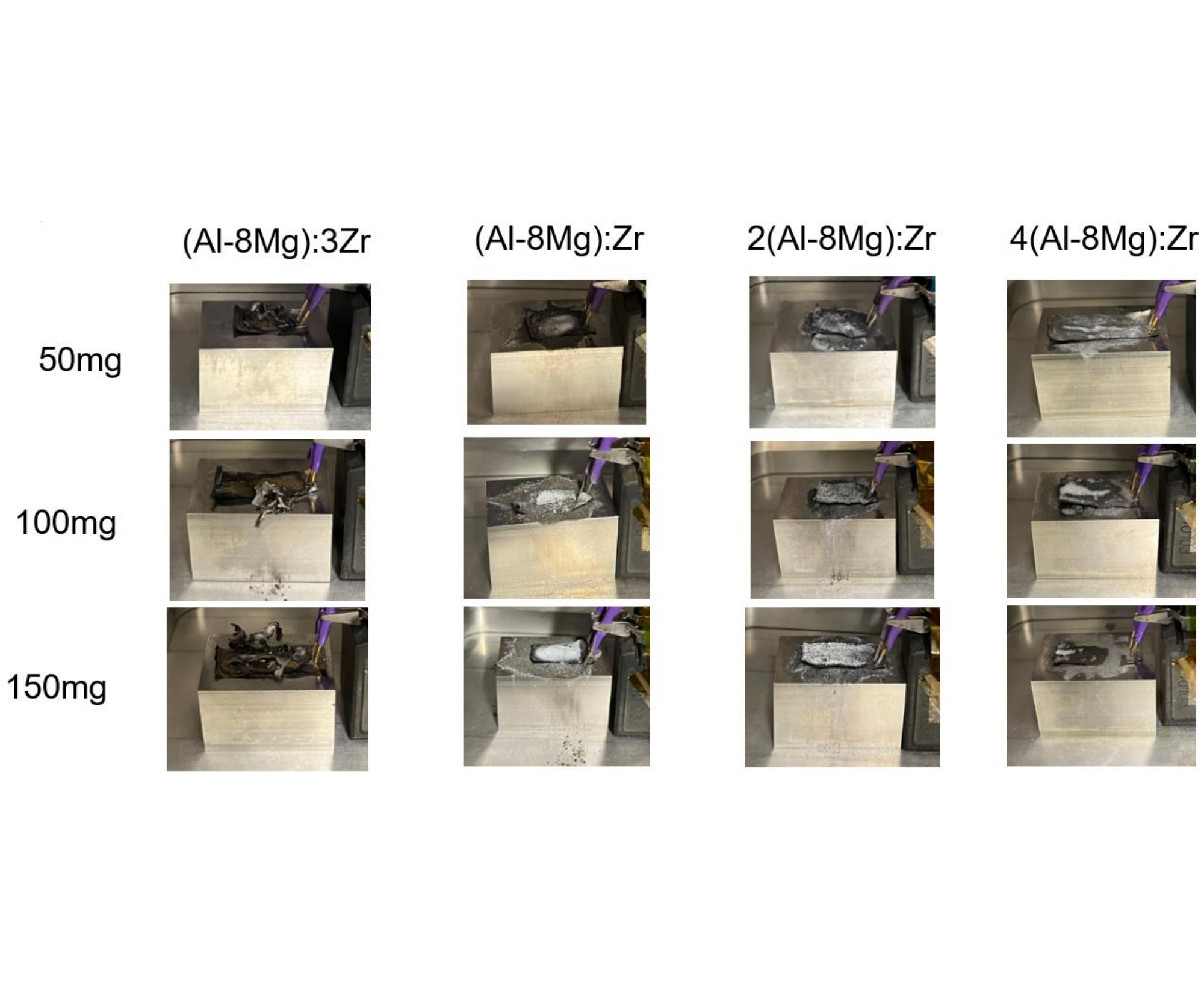}
    \caption{Images of post-fire ignition blocks demonstrating stoichiometric formulations ranging from Zr-rich to Al-rich arranged left to right, and increased loading amounts from top to bottom.}

\end{figure*}
\clearpage

\subsection{Associated Error for Temperature and Defeat Plots}
\setcounter{table}{0}
\renewcommand{\tablename}{Table A}
\begin{table*}[h]
    \centering
    \includegraphics[trim=5cm 5.5cm 5cm 4cm, clip=true,width=0.85\linewidth]{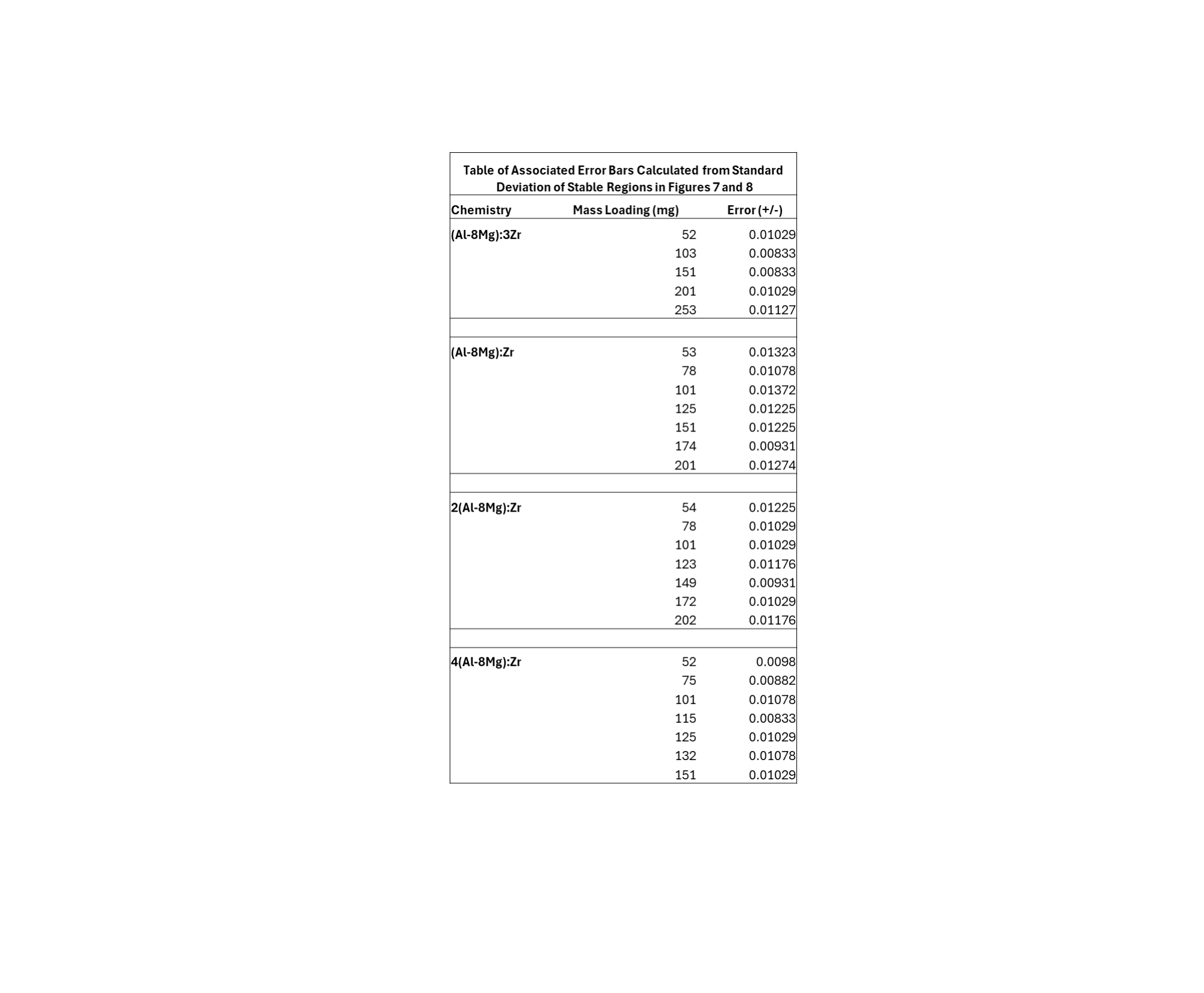}
    \caption{Associated Error for \cref{fig:MassvsDefeat}. Standard deviation is taken from a stable region (~50 data points) and applied over the integrated 5 second timeframe.}
\label{Supplementary3}
\end{table*}

\begin{table*}[h]
    \centering
    \includegraphics[trim=7cm 10.5cm 7cm 9.5cm, clip=true,width=0.85\linewidth]{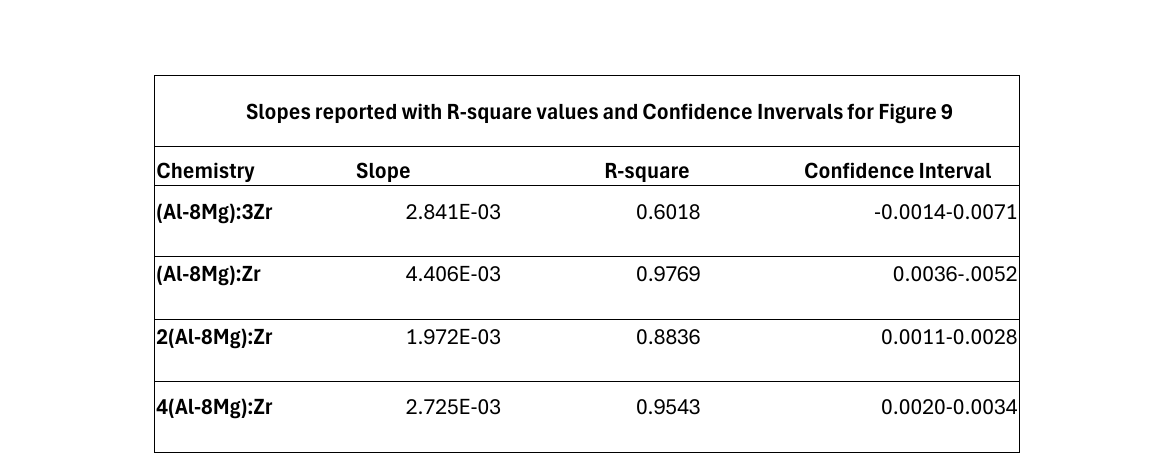}
    \caption{R-square values and confidence intervals for slopes in \cref{fig:TemperaturevsDefeat}}
\label{Supplementary4}
\end{table*}

\clearpage

\section{References}

\twocolumngrid


\bibliography{References}


\end{document}